\documentclass[12pt]{preprint}
\usepackage{Martin}
\usepackage{Symbols}
\usepackage{ScriptFonts}
\usepackage{MHenvs}
\usepackage{times}
\usepackage{amsbsy}
\usepackage[final]{MHequ}

\newcommand\E{\hbox{\bf E}}
\renewcommand\P{\hbox{\bf P}}

\def\B{\cscr{B}}
\def\${|\!|\!|}
\def\law{\cscr L}
\def\qua{\kappa}
\def\mm#1{\langle\!\langle#1\rangle\!\rangle}

\mynumbering
\mydimensions

\begin{document}

\title{Invariant Measures for Stochastic \\[1mm]
PDE's in Unbounded Domains}

\author{J.-P.~Eckmann\inst{1}\fnmsep\inst{2}, M.~Hairer\inst{1}}
\institute{D\'epartement de Physique Th\'eorique, Universit\'e de Gen\`eve \and Section de Math\'ematiques, Universit\'e de Gen\`eve
 \\ \email{Jean-Pierre.Eckmann@physics.unige.ch}\\
\email{Martin.Hairer@physics.unige.ch}}
\titleindent=0.65cm
\maketitle
\thispagestyle{empty}
\pagestyle{MHheadings}
\begin{abstract}
We study stochastically forced semilinear parabolic PDE's of the
Ginzburg-Landau type. The class of forcings considered are white
noises in time and colored smooth noises in space. Existence of the
dynamics in $\L^\infty$, as well as existence of an invariant measure
are proven. We also show that the solutions are with high probability
analytic in a strip around the real axis and give estimates on the
width of that strip. 
\end{abstract}
%
\section{Introduction}
\label{sec:Intro}
%
We consider the stochastic partial differential equation (SPDE) given
by
\begin{equa}[0][e:GL]
\tag{SGL}
du_\xi(t) = \Delta u_\xi(t)\,dt + (1-|u_\xi(t)|^2)u_\xi(t)\,dt + Q\,dW(t)\;,\\
u_\xi(0) = \xi\;,\qquad \xi \in \L^\infty(\R)\;.
\end{equa}
In this equation, $dW(t)$ denotes the canonical cylindrical Wiener process on the Hilbert space $\L^2(\R,dx)$, \ie we have the formal expression 
\begin{equ}
\E\bigl(dW(s,x)\,dW(t,y)\bigr) = \delta(s-t)\delta(x-y)\,ds\,dt\;.
\end{equ}
Think for the moment of $u_\xi(t)$ as a distribution on the real line. We will introduce later the space of functions in which \eref{e:GL} makes sense.
The symbol $Q$ denotes a bounded operator of the type
$Qf = \phi_1 \star (\phi_2\,f)$ where $\hat\phi_1$, the Fourier transform of 
$\phi_1$, is some positive $\cOinf$ function and $\phi_2$ is some smooth function that decays sufficiently fast at infinity to be square-integrable. In fact, we will assume for convenience that there are constants $c>0$ and $\beta > 0$ such that
\begin{equ}[e:condPhi2]
|\phi_2(x)| \le {c \over \mm{x}^{1/2+\beta}}\;,\qquad \mm{x} \equiv \sqrt{1+x^2}\;.
\end{equ}
The space in which we show the existence of the solutions is $C_u(\R)$, the Banach space of complex-valued uniformly continuous functions. The reason of this choice is that we want to work in a translational invariant space which is big enough to contain the interesting part of the dynamics of the deterministic part of the equation, \ie the three fixed points $0$ and $\pm 1$, as well as various kinds of fronts and waves.
The meaning of the assumptions on $\phi_1$ and $\phi_2$ is the following.
\begin{claim}
\item The noise does not shake the solution too badly at infinity (in the space variable $x$). If it did, the solution would not stay in $\L^\infty$.
\item The noise is smooth in $x$ (it is even analytic), so it will not lead to irregular functions in $x$-space. This assumption is crucial for our existence theorem concerning the invariant measure.
\end{claim}
For convenience, we write \eref{e:GL} as
\begin{equa}[0][e:GL2]
du_\xi(t) = \bigl(Lu_\xi(t) + F(u_\xi(t))\bigr)\,dt + Q\,dW(t)\;,\\
L=\Delta-1\;,\quad \bigl(F(u)\bigr)(x) = u(x)+(1-|u(x)|^2)u(x)\;.
\end{equa}
This is also to emphasize that our proofs apply in fact to a much larger class of SPDE's of the form \eref{e:GL2}. For example, all our results apply to the stochastically perturbed Swift-Hohenberg equation
\begin{equ}
du_\xi(t) = (1-\Delta)^2 u_\xi(t)\,dt + (1-|u_\xi(t)|^2)u_\xi(t)\,dt + Q\,dW(t)\;,
\end{equ}
but one has to be more careful in the computations, since one does not know an explicit formula for the kernel of the linear semigroup. It is also possible to replace the nonlinearity by some slightly more complicated expression of $u(t)$.

For any Banach space $\B$, a $\B$-valued stochastic process
$u_{\xi}(t)$ is called a {\it mild solution} of \eref{e:GL2} with initial condition $\xi$ if it satisfies the associated integral equation
\begin{equ}[e:GLInt]
u_{\xi}(t) = e^{L t} \xi + \int_0^t e^{L(t-s)} F\bigl(u_{\xi}(s)\bigr)\,ds + \int_0^t e^{L(t-s)} Q\,dW(s)\;,
\end{equ}
in the sense that every term defines a stochastic process on $\B$ and that the equality holds almost surely with respect to the probability measure on the abstract probability space underlying the Wiener process.
The initial condition does not have to belong to $\B$, provided $e^{Lt}\xi \in \B$ for all times $t>0$.

To a Markovian solution, we can associate (under suitable conditions) the {\it transition semigroup} $\CP_t$ defined on and into the set of bounded Borel functions $\phi : \B \to \C$ by
\begin{equ}[e:defSemi]
\bigl(\CP_t\phi\bigr)(\xi) = \int_{\!\B}\phi(\eta) \P\bigl(u_{\xi}(t)\in d\eta\bigr)\;.
\end{equ}
Its dual semigroup $\CP_t^*$ is defined on and into the set of Borel probability measures $\nu$ on $\B$ by
\begin{equ}[e:defAdj]
\bigl(\CP_t^*\nu\bigr)(\Gamma) = \int_{\!\B}\P\bigl(u_{\xi}(t) \in \Gamma\bigr)\,\nu(d\xi)\;,
\end{equ}
where $\Gamma$ is a $\B$-Borel set. If the existence of the solutions is shown for initial conditions in a larger Banach space $\B'$ in which $\B$ is continuously embedded, $\CP_t^*$ can be extended to a map from the $\B'$-Borel probability measures into the $\B$-Borel probability measures.

An \emph{invariant measure} for \eref{e:GL2} is a probability measure on $\B$ which is a fixed point for $\CP_t^*$. If $\cscr T$ is a weaker topology on $\B$, we can under appropriate conditions extend $\CP_t^*$ by \eref{e:defAdj} to a mapping from the $\cscr T$-Borel probability measures into themselves. In the case of $\L^\infty(\R)$, we may for example consider a ``weighted topology'' $\cscr T_\rho$ induced by some weighted norm $\|\rho \cdot\|_\infty$.

If we take $\phi_2(x) = 1$, it is known (we refer to \cite{ZDP} for details) that \eref{e:GL2} possesses a mild solution in $\L^p(\R,\rho(x)\,dx)$ for a weight function $\rho$ that decays at infinity. Our choice for $Q$ makes it possible to work in flat spaces, since the noise is damped at infinity. In fact, we will show that, for every initial condition $u_0 \in \L^\infty(\R)$, \eref{e:GL2} possesses a mild solution in $C_u(\R)$, the space of bounded uniformly continuous functions on $\R$. This leads to slight technical difficulties since neither $\L^\infty(\R)$ nor $C_u(\R)$ are separable Banach spaces, and thus standard existence theorems do not apply.

After proving the existence of the solutions, we will be concerned with their regularity.
We prove that with high probability the solution
$u_\xi(t)$ of \eref{e:GL} for a fixed time is analytic in a strip
around the real axis.
We will also derive estimates on the width of that strip. These estimates will finally allow to show the existence of an invariant measure for $\CP_t^*$, provided we equip $C_u(\R)$ with a slightly weaker topology. The existence of an invariant measure is not a trivial result since
\begin{claim}
\item[\it a.] The linear semigroup of \eref{e:GL} is not made of compact operators in $C_u(\R)$.
\item[\it b.] The deterministic equation is not strictly dissipative, in the sense that there is not a unique fixed point that attracts every solution.
\item[\it c.] The deterministic equation is of the gradient type, but the operator $Q$ is not invertible, so we can not make the {\it a priori} guess that the invariant measure is some Gibbs measure.
\end{claim}
The results we found in the literature about the existence of invariant measures for infinite-dimen\-sional stochastic differential equations (see \eg \cite{JLM,DZA,ZDP,BC} and references therein) usually assume that the converse of either {\it a.}, {\it b.}~or {\it c.}~holds.
 The main result of this paper is the following.
\begin{theorem}
\label{theo:existM}
There exist slowly decaying weight functions $\rho$ such that the extension of $\CP_t^*$ to the $\cscr T_\rho$-Borel probability measures is well-defined and admits a fixed point.
\end{theorem}
\begin{remark}
The hypotheses of this theorem have been made with the following future
project in mind. We hope to prove that the measure found in \theo{theo:existM} is \emph{unique}. The basic idea is to apply the methods
of \cite{EPR2} to the context of SPDE's to show uniqueness of the
measure by the tools of control theory. In this context,
it is interesting if the noise drives the system only in the
dissipative range, namely in a \emph{finite} interval of frequencies
which need not contain the unstable modes of the deterministic Ginzburg-Landau equation. In particular, such forces do \emph{not} have invertible
covariances and hence methods such as those found in \cite{ZDP} do not apply.

This is also the reason why the setting considered in this paper imposes $\hat\phi_1$ to have compact support, although the extension to exponentially decaying functions would have been easy.
\end{remark}
The next sections will be organized as follows. In \sect{sec:Conv}, we give detailed bounds on the stochastic convolution, \ie on the evolution of the noise under the action of the semigroup generated by $L$. In \sect{sec:exist} we then prove the existence of a unique solution for \eref{e:GL2} and derive an {\it a priori} estimate on its amplitude. \sect{sec:anal} is devoted to the study of the analyticity properties of the solution. In \sect{sec:inv}, we finally show the existence of an invariant measure for the dynamics, \ie we prove \theo{theo:existM} which will be restated as \theo{theo:exist}. The appendix gives conditions under which one can prove the existence of a global strong solution to a class of semilinear PDE's in a Banach space.
%
\subsection{Definitions and notations}
\label{sec:Defs}
%
Consider the sets $\cscr A_\eta$ of functions that are analytic and uniformly bounded in
an open strip of width $2\eta$ centered around the real axis. They are Banach spaces with respect to the norms
\begin{equ}
\$f\$_{\eta,\infty} \equiv \sup_{z \,:\, |\Im z| < \eta} |f(z)|\;.
\end{equ} 
Fix $T > 0$. We define $\B_T$ as the Banach space of functions
$f(t,x)$ with $t \in (0,T]$ and $x \in \R$ such that for fixed $t>0$, 
$f(t,\cdot)$ is analytic and bounded in the strip $\{z=x+iy\,|\,|y| < \sqrt{t}\}$.
We equip $\B_T$ with the norm
\begin{equ}
\$f\$_{T} \equiv \sup_{t \in (0,T]} \$f(t,z)\$_{\sqrt t,\infty}\;.
\end{equ}
In the sequel we denote by $\|\cdot\|_{p}$ the norm
of $\L^p(\R,dx)$. For $M$ a metric space and $\B$ a Banach space, the symbol $C_b(M,\B)$ (resp.~$C_u(M,\B)$) stands for the Banach space of bounded (uniformly) continuous functions $M \to \B$ endowed with the usual sup norm. If $\B = \C$, it is usually suppressed in the notation.
Moreover, the symbol $C$ denotes
a constant which is independent of the running parameters and which may change from one line to the other (even inside the same equation).

The symbol $\law(X)$ denotes the probability law of a random variable $X$. The symbol $\CB(M,r)$ denotes the open ball of radius $r$ centered at the origin of a metric vector space $M$.
%
\section{The Stochastic Convolution}
\label{sec:Conv}
%
This section is devoted to the detailed study of the properties
of the stochastic process obtained by letting the semigroup generated by $L$ act on the noise.
%
\subsection{Basic properties}
\label{sec:Prop}
%
Let us denote by $(\Omega,\cscr F,\P)$ the underlying probability space
for the cylindrical Wiener process $dW$, and by $\E$ the expectation in $\Omega$.
We define the stochastic convolution
\begin{equ}[e:defStochWL]
W_L(t,\omega) = \int_0^t e^{L(t-s)}Q\,dW(s,\omega)\;,\quad \omega \in \Omega\;.
\end{equ}
The argument $\omega$ will be suppressed during the major part of the discussion.
For a discussion on the definition of the stochastic integral in
infinite-dimensional Banach spaces, we refer to \cite{ZDP1}.
Notice that since $\hat\phi_1$ has compact support, we can find
a $\cOinf$ function $\hat\psi$ such that $\hat\psi(x)=1$
for $x\in\supp\hat\phi$. We define $\tilde Q f = \psi \star f$
and fix a constant $R$ such that
\begin{equ}[e:defR]
\supp\hat\phi \subset \supp \hat\psi \subset \{x \in \R \,|\, |x| \le R\}\;.
\end{equ}
We have of course $\tilde Q Q = Q$.
An important consequence of this property is
\begin{lemma}
\label{lem:boundW}
Fix $\eta > 0$ and $\alpha < 1/2$. Then there exists a version
of $W_L$ with $\alpha$-H\"older continuous sample paths in $\cscr A_{\eta}$. Furthermore, for every $T>0$, the mapping
\begin{equa}[e:defWL]
W_L^\eta : \Omega &\to C_b([0,T],\cscr A_{\eta})\;,\\
\omega &\mapsto W_L(\cdot,\omega)\;,
\end{equa}
is measurable with respect to the Borel $\sigma$-field generated by the strong topology on $C_b([0,T],\cscr A_{\eta})$.
\end{lemma}
\begin{remark}
The meaning of the word ``version'' is that the process constructed here differs from \eref{e:defStochWL} only on a set of $\P$-measure $0$. We will in the sequel not make any distinction between both processes.
\end{remark}
\begin{proof}[of \lem{lem:boundW}]
We first notice that $W_L(t)$ has an $\alpha$-H\"older continuous version in $\L^2(\R)$. This
is a consequence of the fact that the Hilbert-Schmidt norm in $\L^2(\R)$ of $\exp(Lt)Q$ is bounded by $e^{-t}\|\phi_1\|_2\|\phi_2\|_2$. Since $\L^2(\R)$ is separable, the mapping
\begin{equs}
W_L : \Omega &\to C_b([0,T], \L^2(\R))\;,\\
\omega &\mapsto W_L(\cdot,\omega)\;,
\end{equs}
is measurable \cite[Prop~3.17]{ZDP1}.
Since $L$ and $\tilde Q$ commute, we can write
\begin{equ}[e:altW]
W_L(t,\omega) = \int_0^t  \tilde Q^2 e^{L(t-s)} Q\,dW(s,\omega) = \tilde Q^2 W_L(t,\omega)\;,
\end{equ}
where we used \cite[Prop.~4.15]{ZDP} to commute the operator and the integral.
We will show that $\tilde Q^2$ defines a bounded continuous linear operator from $\L^2(\R)$ into $\cscr A_{\eta}$. The claim then follows if we define the map $W_L^\eta = \tilde Q_\eta^2 \circ W_L$, where we denote by $\tilde Q^2_\eta$ the operator constructed in an obvious way from $\tilde Q^2$ as a map from $C_b([0,T], \L^2(\R))$ into $C_b([0,T], \cscr A_\eta)$.

Notice first that if $f \in \L^2(\R)$, we have by the Young inequality $\tilde Q f \in \L^\infty(\R)$ and the estimate
\begin{equ}[e:esttildeQ1]
\|\tilde Q f\|_\infty \le \|\psi\|_2 \|f\|_2
\end{equ}
holds.
Take now $f \in \L^\infty(\R)$. Since $\tilde Q$ maps any measurable function onto an entire analytic function, $\tilde Q f(z)$ has a meaning for every $z \in \C$. We have for any $x\in \R$
\begin{equ}[e:estWrho]
\bigl|\bigl(\tilde Qf\bigr)(x+i\eta)\bigr| = \Bigl|\int_R \psi(x+i\eta-y) f(y)\,dy\Bigr|\;.
\end{equ}
By assumption, the Fourier transform of $\psi$ belongs to $\cOinf$. 
We know that such functions enjoy the
property -- see \eg \cite{RS} -- that for each $N>0$ there exists a constant $C_N$ such that
\begin{equ}
|\psi(x+i\eta)| \le {C_N e^{R|\eta|} \over (1+x^2 + \eta^2)^N}\;,
\end{equ}
where the constant $R$ is defined in \eref{e:defR}.
We thus have the estimate
\begin{equa}[e:esttildeQ2]
\bigl|\bigl(\tilde Qf\bigr)(x+i\eta)\bigr| &\le \|f\|_{\infty}\int_\R |\psi(x+i\eta-y)|\,dy \\
&\le C e^{R|\eta|} \|f\|_{\infty}\;,
\end{equa}
and thus
\begin{equ}[e:estBanach]
\$\tilde Qf\$_{\eta,\infty} \le C e^{R|\eta|} \|f\|_{\infty}\;.
\end{equ} 
Collecting \eref{e:esttildeQ1} and \eref{e:estBanach} proves the claim.
\end{proof}
\begin{remark}
\label{rem:boundW}
As an evident corollary of the proof of the lemma, note that $W_L(t) \in \CD(L)$ for all times $t\ge 0$ and that the mapping
\begin{equa}[e:defWLL]
W_L : \Omega &\to C_b([0,T], \CD(L))\;,\\
\omega &\mapsto W_L(\cdot,\omega)\;,
\end{equa}
has the same properties as the mapping $W_L^\eta$ if we equip $\CD(L)$ with the graph norm. In particular, $W_L$ has almost surely $\alpha$-H\"older continuous sample paths in $\CD(L)$.
\end{remark}
We will now give more precise bounds on the magnitude of the process $W_L$. Our main tool will be the so-called ``factorization formula'' which will allow to get uniform bounds over some finite time interval.
%
%
\subsection{Factorization of the stochastic convolution}
\label{sec:Fact}
%
%
We define, for $\delta \in (0,1/2)$,
\begin{equs}
Y_{L,\delta}(t) &= \int_0^t (t-s)^{-\delta}e^{L(t-s)}Q\,dW(s)\;,\\
\bigl(G_{\delta}\Psi\bigr)(t) &= \int_0^t (t-s)^{\delta-1}e^{L(t-s)}\Psi(s)\,ds\;.
\end{equs}
Notice that we can show by the same arguments as in \lem{lem:boundW} that the process $Y_{L,\delta}(t)$ has a version which takes values in $\cscr A_\eta$. Thus, in particular the expression $Y_{L,\delta}(t,x)$ is a well-defined complex-valued random variable.
A corollary of the stochastic Fubini theorem (sometimes referred to as the
``factorization formula'' \cite{ZDP1}) shows that
\begin{equ}[e:Fact]
W_L(t) = {\sin \pi\delta \over \pi}\bigl(G_{\delta}Y_{L,\delta}\bigr)(t)\;.
\end{equ}
Before we start to estimate $\|W_L(t)\|_{\infty}$, we state without proof the
following trivial consequence of the Young inequality:
\begin{lemma}
\label{lem:estKernel}
Denote by $g_t$ the heat kernel and choose $p > 1$. Then there exists a constant $c$ depending on $p$ such that
\begin{equ}[e:estKernel]
\|g_t \star f\|_{\infty} \le ct^{-1/(2p)}\|f\|_p\;,
\end{equ}
holds for every $f \in \L^p(\R)$.
\end{lemma}
We have, using \eref{e:Fact}, \lem{lem:estKernel}, and the H\"older inequality,
\begin{equs}
\|W_L(t)\|_{\infty} &\le C \int_0^t (t-s)^{\delta-1}e^{-(t-s)}\|g_{t-s}\star 
	Y_{L,\delta}(s)\|_{\infty}\,ds \\
	&\le C \int_0^t (t-s)^{\delta-1-1/(2p)}\|Y_{L,\delta}(s)\|_p\,ds\\
&\le C \Bigl(\int_0^t (t-s)^{q(\delta-1-1/(2p))}\,ds\Bigr)^{1/q}
\Bigl(\int_0^t \|Y_{L,\delta}(s)\|_p^p\,ds\Bigr)^{1/p}\;,
\end{equs}
where $q$ is chosen such that $p^{-1} + q^{-1} = 1$.
It is easy to check that the first integral converges when
\begin{equ}[e:condP]
p > {3\over 2\delta}\;.
\end{equ}
In that case, we have
\begin{equ}[e:estW]
\|W_L(t)\|_{\infty}^p \le C t^\gamma\int_0^t \|Y_{L,\delta}(s)\|_p^p\,ds\;,\qquad \gamma = p\delta - {3\over 2}\;.
\end{equ}
So it remains to estimate $\|Y_{L,\delta}(t)\|_p$. 

%
\subsection{Estimate on the process $\boldsymbol{Y_{L,\delta}(t)}$}
%

This subsection is devoted to the proof of the following lemma.
\begin{lemma}
\label{lem:estY}
Let $Y_{L,\delta}$ be as above and choose $p \ge 2$ and $\delta \in (0,1/2)$. There exists a constant $c$ depending on $\delta$, $p$, $\phi_1$ and $\phi_2$ but independent of $t$ such that $\E \|Y_{L,\delta}(t)\|_{p}^p \le c$.
\end{lemma}
Remember that the convolution of two decaying functions decays like
the one that decays slower at infinity:
\begin{lemma}
\label{lem:estConv}
Let $f$ and $g$ be two positive even functions which are integrable and monotone decreasing between $0$ and $\infty$. Then the estimate
\begin{equ}
|(f\star g)(x)| \le |f(x/2)|\, \|g\|_1 + |g(x/2)|\, \|f\|_1
\end{equ}
holds.
\end{lemma}
\begin{proof}
Assume $x \ge 0$ (the case $x < 0$ can be treated in a similar way) and define $I_x = (x/2,3x/2)$. We can decompose the convolution as
\begin{equs}
|(f\star g)(x)| &\le \int_{I_x} |f(y-x)g(y)|\,dy + \int_{\R \setminus I_x} |f(y-x)g(y)|\,dy\\
&\le |g(x/2)|\int_{\R} |f(y)|\,dy + |f(x/2)|\int_{\R} |g(y)|\,dy\;,
\end{equs}
which proves the assertion.
\end{proof}
\begin{proof}[of \lem{lem:estY}]
We use the formal expansion
\begin{equ}
dW(x,t) = \sum_{j=1}^\infty e_j(x)\,dw_j(t)\;,
\end{equ}
where
the $e_i$ form an orthonormal basis of $\L^2(\R,dx)$ (say the eigenfunctions of the harmonic oscillator) and the $dw_i$ are independent
Wiener increments. We also denote by $T_x$ the translation operator $(T_x f)(y) = f(y-x)$. We then have
\begin{equs}
\E|Y_{L,\delta}(t,x)|^2 &= \E \biggl|\int_0^t\sum_{j=1}^\infty (t-s)^{-\delta} e^{-(t-s)} (g_{t-s} \star \phi_1 \star (\phi_2\,e_j))(x)\,dw_j(s)\biggr|^2 \\
&= \int_0^t \sum_{j=1}^\infty (t-s)^{-2\delta} e^{-2(t-s)} |(g_{t-s} \star \phi_1 \star (\phi_2\,e_j))(x)|^2\,ds \\
&= \int_0^t (t-s)^{-2\delta} e^{-2(t-s)} \sum_{j=1}^\infty \bigl|\scal{\phi_2\,T_x(g_{t-s} \star \phi_1), e_j}\bigr|^2\,ds \\
&= \int_0^t s^{-2\delta} e^{-2s} \|\phi_2\,T_x(g_{s} \star \phi_1)\|_2^2 \,ds\;.
\end{equs}
An explicit computation shows the equality
\begin{equ}
\|\phi_2\,T_x(g_{s} \star \phi_1)\|_2^2 = \bigl(\phi_2^2\star(g_s \star \phi_1)^2\bigr)(x)\;.
\end{equ}
Using \lem{lem:estConv}, the fact that $\phi_1(x) \le C_N \mm{x}^{-N}$ for every $N$, and the well-known inequality $|g_s \star \phi_1|(x) \le \|\phi_1\|_\infty$, we get the estimate
\begin{equ}
(g_s \star \phi_1)^2(x) \le C\Bigl({e^{-x^2/(16s)}\over \mm{s}} + {1 \over \mm{x}^{N}}\Bigr)\;.
\end{equ}
Using again \lem{lem:estConv} and \eref{e:condPhi2}, we get
\begin{equ}
\|\phi_2\,T_x(g_{s} \star \phi_1)\|_2^2 \le C\Bigl({e^{-x^2/(64s)}\over \mm{s}} + {1 \over \mm{x}^{1+2\beta}}\Bigr)\;.
\end{equ}
It is now an easy exercise to show that
\begin{equ}
\sup_{s > 0} \|\phi_2\,T_x(g_{s} \star \phi_1)\|_2^2 \le C\Bigl({1 \over \mm{x}^{2}} + {1 \over \mm{x}^{1+2\beta}}\Bigr)\;.
\end{equ}
Defining $\beta' = \min\{1/2,\beta\}$, and using $\mm{x} \ge 1$, we have
\begin{equ}
\E|Y_{L,\delta}(t,x)|^2 \le C \mm{x}^{-1-2\beta'} \int_0^t s^{-2\delta} e^{-2s}\,ds \le
	C \mm{x}^{-1-2\beta'}\;.
\end{equ}
Since $Y_{L,\delta}(t,x)$ is a Gaussian random variable, this implies, for $p \ge 2$
\begin{equa}[e:estY]
\E \|Y_{L,\delta}(t)\|_{p}^p &= \int_{\R}\E |Y_{L,\delta}(t,x)|^p\,dx \le C \int_{\R}\bigl(\E |Y_{L,\delta}(t,x)|^2\bigr)^{p/2}\,dx \\
&\le C \int_{\R} {1 \over \mm{x}^{p/2 + \beta' p}}\,dx \le C\;.
\end{equa}
This proves the assertion.
\end{proof}
As a corollary of \lem{lem:estY}, we have the following estimate on the process $W_L(t)$.
\begin{corollary}
\label{cor:estW}
For any $p\ge 2$, there is a constant $C>0$ such that $\E\|W_L(t)\|_\infty^p \le C$ for all times $t\ge 0$.
\end{corollary}
\begin{proof}
Using again the equality $W_L(t) = \tilde Q W_L(t)$, we notice that it
is enough to have an estimate on $\E\|W_L(t)\|_p^p$. This can be done by retracing the proof of \lem{lem:estY} with $\delta$ replaced by $0$.
\end{proof}
We have now collected all the necessary tools to obtain the main result of this section.
\begin{theorem}
\label{theo:Main}
For every $\eps > 0$, there are constants $C,R>0$ depending only on the choices of $\phi_1$, $\phi_2$ and $\eps$ such that the estimate
\begin{equ}
\E\$W_L\$_{T} \le C e^{R\sqrt T}T^{1/2 - \eps}
\end{equ}
holds. \qed
\end{theorem}
\begin{proof}
The estimate
\begin{equ}[e:estWanal]
\$W_L\$_{T} \le C e^{R\sqrt T} \sup_{t \in (0,T]} \|W_L(t)\|_{\infty}\;,
\end{equ}
holds as a consequence of Eqs.~\eref{e:estWrho} and \eref{e:estBanach}.
We thus need an estimate on $\|W_L(t)\|_{\infty}$ which is uniform on some time interval. This is achieved by combining
\lem{lem:estY} with Eq.~\eref{e:estW}. Let us first choose a constant $\delta>1/2$, but very close to $1/2$ and then a (big) constant $p$ such that $p > \max\{2,3/(2\delta)\}$. Since
$\sup_{t \in (0,T]} \|W_L(t)\|_{\infty}$ is a positive random variable, we have
\begin{equs}
\E \Bigl(\,\displaystyle{\sup_{t \in (0,T]}} \|W_L(t)\|_{\infty}\Bigr)
&\le C \Bigl(\E\bigl(\textstyle{\sup_{t \in (0,T]}} \|W_L(t)\|_{\infty}\bigr)^p\Bigr)^{1/p} \\
&=  C \Bigl(\E\bigl(\textstyle{\sup_{t \in (0,T]}} \|W_L(t)\|_{\infty}^p\bigr)\Bigr)^{1/p}\\
& \le C\Bigl(T^\gamma\int_0^T \E \|Y_{L,\delta}(s)\|_{p}^p\,ds\Bigr)^{1/p} \\
&\le C T^{(\gamma + 1)/p} \le C T^{\delta - 1/(2p)}\;.\label{e:estsupW}
\end{equs}
The exponent $\delta - 1/(2p)$ can be brought arbitrarily close to $1/2$.
This, together with the previous estimate \eref{e:estWanal}, proves the claim.
\end{proof}
We have now the necessary tools to prove the existence of a unique solution to the SPDE \eref{e:GL2}.
%
%
%
\section{Existence of the Solutions}
\label{sec:exist}
%
%
%
Throughout this section, we denote by $\B$ the Banach space $C_u(\R)$ of bounded uniformly continuous complex-valued functions on the real line endowed with the norm $\|\cdot\|_\infty$. The reason why we can not use a standard existence theorem is that $\B$ is not separable. Nevertheless, the outline of our proof is quite similar to the proofs one can find in \cite{ZDP1}. The technique is to solve \eref{e:GL2} pathwise and then to show that the result yields a well-defined stochastic process on $\B$ which is a mild solution to the considered problem. In order to prepare the existence proof for solutions of \eref{e:GL2}, we study the
dynamics of the \emph{deterministic} equation
\begin{equ}[e:path]
\dot X_\xi(W,t) = LX_\xi(W,t) + F(X_\xi(W,t) + W(t)) \;, \quad X_\xi(W,0) = \xi.
\end{equ}
In this equation, $\xi \in \L^\infty(\R)$ is an arbitrary initial condition and $W \in C_b([0,T],\cscr A_\eta)$ is an arbitrary noise function with $W(0) = 0$ and $\eta>0$ fixed. For the moment, we choose an arbitrary time $T>0$ and study the solutions up to time $T$. The reason why we study \eref{e:path} is that if $X_\xi$ is a solution of \eref{e:path}, then $Y_\xi(t) = X_\xi(t) + W(t)$ is a solution of
\begin{equ}
\dot Y_\xi(t) = LY_\xi(t) + F(Y_\xi(t)) + \dot W(t) \;, \quad Y_\xi(0) = \xi,
\end{equ}
provided $W:[0,T]\to \cscr A_\eta$ is a differentiable function.
Because of the dissipativity of $F$, we will show that \eref{e:path} possesses a unique bounded and continuous solution in $\B$ for all times $t \in (0,T]$. Consider the map
\begin{equa}
 S^T_\xi : C_b([0,T], \cscr A_\eta) &\to C_b((0,T],\B)\;, \\
	W(\cdot)\qquad &\mapsto X_\xi(W,\cdot)\;,
\end{equa}
that associates to every noise function $W$ and every initial condition $\xi \in\L^\infty(\R)$ the solution of \eref{e:path}. (We do not show explicitly the value of $\eta$ in the notations, since the map $S^T_\xi$ is in an obvious sense independent of $\eta$.) We have the following result.
\begin{lemma}
\label{lem:detequ}
The map $(\xi,W) \mapsto S^T_\xi(W)$ is locally Lipschitz continuous in both arguments. Furthermore, the estimates
\minilab{e:est}
\begin{equs}
\|S^T_{\xi}(W)\| &\le \max\bigl\{\|\xi\|_\infty,C(1+\|W\|^3)\bigr\}\;,\label{e:est1}\\
\|S^T_\xi(W) - S^T_\zeta(W)\| &\le e^T \|\xi - \zeta\|_\infty\;, \label{e:est2}
\end{equs}
hold.
\end{lemma}
\begin{proof}
The proof relies on the results of \app{app:Diss}.
As a first step, we verify that the assumptions of \theo{theo:Dissip} are satisfied with $F_t(x) = F(x+W(t))$. It is well-known \cite{Lun} that \ass{1} is satisfied for the Laplacean and thus for $L$. Using the easy-to-check inequality
\begin{equ}
|(a-b)+\alpha (a|a|^2-b|b|^2)| \ge |a-b|\Bigl(1+\alpha {|a|^2 + |b|^2 \over 2}\Bigr)\;,
\end{equ}
which holds for any $a,b \in \C$ and $\alpha \ge 0$,
it is also straightforward to check that the mapping $L + F_t$ is $\qua$-quasi dissipative for all times with $\qua = 1$ and therefore \ass{2} holds.
Assumption \ass{3} can be checked in a similar way. To check \ass{4}, notice that by Cauchy's integral representation theorem, $\cscr A_\eta \subset \CD(L)$, and so $F_t$ maps $\CD(L)$ into itself. Furthermore, it is easy to check the inequality
\begin{equ}[e:estDer]
\|\d_x v\|^2_\infty \le C\|v\|_\infty \|\d_x^2 v\|_\infty\;,\qquad v \in \CD(L)\;.
\end{equ}
We leave it to the reader to verify, with the help of \eref{e:estDer}, that \ass{4} is indeed satisfied.
It is clear by the continuity of $W(\cdot)$ that \ass{5} holds as well, so we are allowed to use \theo{theo:Dissip}.

We will show that \eref{e:est} holds for arbitrary initial conditions in $\CD(L)$. To show that they also hold for arbitrary initial conditions in $\L^\infty(\R)$, we can apply arguments similar to what is done at the end of the proof of \theo{theo:Dissip}.

Until the end of the proof, we will always omit the subscript $\infty$ in the norms. Denote by $X(t)$ the solution of \eref{e:path}. Since $X(t)$ is strongly differentiable by \theo{theo:Dissip}, the left lower Dini derivative $D_-\|X(t)\|$ satisfies by \eref{e:Dini}
\begin{equa}[e:subdiff]
D_- \|X(t)\| &\le \liminf_{h\to 0^+} h^{-1}\bigl(\|X(t)\| - \|X(t) - h LX(t) - hF_t(X(t))\|\bigr)\\
&\le -\|X(t)\| + C\bigl(1+\|W(t)\|^3\bigr)\;,
\end{equa}
where the last inequality is easily obtained by inspection, absorbing the linear instability into the strongly dissipative term $-X(t)|X(t)+W(t)|^2$. The estimate \eref{e:est1} follows immediately from a standard theorem about differential inequalities \cite{Wa}.

Inequality \eref{e:est2} is an immediate consequence of \theo{theo:Dissip}.

It remains to show that $S_\xi^T(W)$ is a locally Lipschitz continuous function of $W$. We call $X(t)$ and $\tilde X(t)$ the solutions of \eref{e:path} with noise functions $W$ and $V$ respectively. We also denote by $F_t^W$ and $F_t^V$ the corresponding nonlinearities. In a similar way as above, we obtain the inequality
\begin{equs}
D_- \|X(t) - \tilde X(t)\| &\le \|X(t) -\tilde X(t)\| + {\|(F_t^W-F_t^V)(X(t))\| \over 2} \\
&\qquad + {\|(F_t^W-F_t^V)(\tilde X(t))\| \over 2}\;.
\end{equs}
The claim now follows from the estimate
\begin{equ}
\|(F_t^W-F_t^V)(x)\| \le C\|W-V\|\bigl(1+\|x\|^2 + \|W\|^2 + \|V\|^2\bigr)\;,
\end{equ}
and from the {\it a priori} estimate \eref{e:est1} on the norms of $X(t)$ and $\tilde X(t)$.
\end{proof}
Before we state the existence theorem, let us define the following.
\begin{definition}
A transition semigroup $\CP_t$ on a Banach space $\B$ has the \emph{weak Feller} property if $\CP_t\phi\in C_u(\B)$ for every $\phi \in C_u(\B)$.
\end{definition}
\begin{theorem}
\label{theo:sol}
For every initial condition in $\L^\infty(\R)$, the SPDE defined by \eref{e:GL} possesses a unique continuous mild solution
in $\B$ for all times. The solution is Markov, its transition semigroup is well-defined and weak Feller and its sample paths are almost surely $\alpha$-H\"older continuous for every $\alpha< 1/2$.
\end{theorem}
\begin{proof}
The main work for the proof was done in \lem{lem:detequ}. Recall the definition \eref{e:defWL} of the mapping $W_L^\eta$ that associates to every element of $\Omega$ a continuous noise function in $\cscr A_\eta$. Since $\cscr A_\eta$ is continuously embedded in $\B$, we can \emph{define} the random variable
\begin{equa}
u^T_\xi : \Omega &\to C_b((0,T],\B)\;, \\
	\omega &\mapsto \bigl(S^T_\xi \circ W_L^\eta\bigr)(\omega) + W_L^\eta(\omega)\;,
\end{equa}
for some $\eta > 0$ and some $T > 0$. This allows to define the stochastic process
\begin{equa}
u_\xi(t) : \Omega &\to \B\;, \\
	\omega &\mapsto \bigl(u^T_\xi(\omega)\bigr)(t)\;,
\end{equa}
for some $T>t$.
It is clear by the uniqueness of the solutions to the deterministic equation \eref{e:path} that this expression is well-defined, \ie does not depend on the parti\-cular choice of $T$. It is also independent of the choice of $\eta$. Since $W_L^\eta$ is measurable and $S^T_\xi$ is continuous, $u_\xi$ is a well-defined stochastic process with values in $\B$. It is immediate from the definitions of $W_L^\eta$ and $S^T_\xi$ that $u_\xi$ is indeed a mild solution to \eref{e:GL}. The Markov property follows from the construction and the Markov property of $W_L$.

To show that the transition semigroup is well-defined, it suffices by Fubini's theorem to show that the function
\begin{equ}
P_{\xi,t}(\Gamma) = \P(u_\xi(t) \in \Gamma) = \int_\Omega \chi_\Gamma\bigl(u_\xi(t,\omega)\bigr)\,\P(d\omega)\;,
\end{equ}
is measurable as a function of $\xi$ for every $\B$-Borel set $\Gamma$ and every $t \ge 0$. This is (again by Fubini's theorem) an immediate consequence of the measurability of $W_L$ and the joint continuity of $S_\xi^\eta(W)$.

The weak Feller property is an immediate consequence of \eref{e:est2}, since
\begin{equ}
\bigl|\bigl(\CP_t \phi)(\xi) - \bigl(\CP_t \phi)(\zeta)\bigr|
\le \int_{\!\Omega} \bigl|\phi\bigl(u_\xi(t,\omega)\bigr)-\phi\bigl(u_\zeta(t,\omega)\bigr)\bigr|\,\P(d\omega)\;.
\end{equ}
Now choose $\eps > 0$. Since $\phi \in C_u(\B)$, there exists $\delta >0$ such that $|\phi(x)-\phi(y)|<\eps$ for $\|x-y\| < \delta$. It suffices to choose $\xi$ close enough to $\zeta$ such that $\|u_\xi(t,\omega) - u_\zeta(t,\omega)\| \le e^t\|\xi-\zeta\|< \delta$ holds.

The $\alpha$-H\"older continuity of the sample paths is a consequence of the strong differentiability (and thus local Lipschitz continuity) of the solutions of \eref{e:path} and of the almost sure $\alpha$-H\"older continuity of the sample paths of $W_L$.
\end{proof}
We now show that the solution of \eref{e:GL2} not only exists in $C_b(\R)$ but also stays boun\-ded in probability. In fact we have
\begin{lemma}
\label{lem:estu}
Let $u_\xi(t)$ be the solution of \eref{e:GL2} constructed above with $\xi \in \L^\infty(\R)$. There exist a time $T^* > 0$ depending on $\xi$ and a constant $C>0$ such that $\E \|u(t)\|_\infty \le C$ for every time $t > T^*$.
\end{lemma}
\begin{proof}
From \eref{e:subdiff}, we obtain the estimate
\begin{equ}
\|u(t) - W_L(t)\|_\infty \le e^{-t} \|\xi\|_\infty + C\int_0^t e^{-(t-s)}\bigl(1 + \|W_L(s)\|_\infty\bigr)^3\, ds\;.
\end{equ}
This yields immediately
\begin{equ}
\sup_{t > T}\E \|u(t)\|_\infty \le e^{-T} \|\xi\|_\infty + C\sup_{s > 0}\E \bigl(1 + \|W_L(s)\|_\infty + \|W_L(s)\|_\infty^3\bigr)\;.
\end{equ}
The claim follows now easily from \cor{cor:estW}.
\end{proof}
%
%
%
\section{Analyticity of the Solutions}
\label{sec:anal}
%
%
%
Our first step towards the existence proof for an invariant measure consists in proving that the solution of \eref{e:GL} constructed in \sect{sec:exist} lies for
 all times in some suitable space of analytic functions. More precisely, we show that there is a (small) time $T$ such that the solution of \eref{e:GL} up to time $T$ belongs to $\B_T$. (Recall the definition of $\B_T$ given in Subsection~\ref{sec:Defs}.) The 
proof is inspired by that of \cite{PC} for the deterministic case, making use of the estimates of
 the preceding
sections, in particular of \theo{theo:Main}.

We split the evolution into a linear part and the remaining
nonlinearity. Recall the definitions
\begin{equ}
L = \Delta - 1\quad\text{and}\quad F(u)(x) = u(x)\bigl(2-|u(x)|^2\bigr)\;.
\end{equ}
Throughout this section, we assume that $u(t)$ is a stochastic process
solving \eref{e:GL} in the mild sense, \ie  there exists a $\xi \in \L^\infty(\R)$ such that $u(t)$ satisfies \eref{e:GLInt}. Such a process exists and is unique (given $\xi$) by \theo{theo:sol}.

For given functions $g \in \L^\infty(\R)$ and $h \in \B_T$,
we define the map $\CM_{g,h} : \B_T \to \B_T$ as
\begin{equa}[e:defM]
\bigl(\CM_{g,h}(f)\bigr)(t) &= h(t) + e^{Lt}g
+ \int_0^t e^{L\tau}F\bigl(f(t-\tau)\bigr)\,d\tau \\
&\equiv h(t) + (\CL g)(t) + (\CN f)(t)\;.
\end{equa}
Until the end of this proof, we write $\$\cdot\$$ instead of $\$\cdot\$_T$. It is possible to show -- see \cite{PC} -- that $\CM_{g,h}$ is always well-defined on $\B_T$ and that
 there are constants $k_1$, $k_2$, $k_3$ such that
\begin{equs}
\$\CL g\$ &\le k_1 \|g\|_\infty\;,\\
\$\CN f\$ &\le k_2 T \$f\$^3\;,\\
\$\CM_{g,h} f_1 - \CM_{g,h} f_2\$ &\le k_3 T (1 + \$f_1\$ + \$f_2\$)^2\$f_1 - f_2\$\;.
\end{equs}
We now show that $u(t) \in \cscr A_\eta$ with
high probability for some $\eta > 0$. The precise statement of the result is
\begin{theorem}
\label{theo:estAnal}
For any $\eps > 0$ there are constants $\eta, \tilde T, C > 0$ such that
$\P(u(t) \in \cscr \CB(\cscr A_\eta,C)) > 1-\eps$ for every time $t > \tilde T$.
\end{theorem}
\begin{proof}
We fix $\tilde T$ bigger than the value $T^*$ we found in \lem{lem:estu}, say $\tilde T=T^*+1$. We also fix some time $T <1$ to be chosen later and we choose an arbitrary time $t>\tilde T$.
We show that with high probability, the solution $u(t-T+\cdot)$
belongs to $\B_T$.
To begin, we take $g = u(t-T)$ and, for $s>0$, we define
\begin{equ}
h(s) = \int_{t-T}^{t-T+s} \!e^{L(t-T+s - \sigma)}Q\,dW(\sigma)\;.
\end{equ}
Since the Wiener increments are identically distributed independent random variables, it is clear that $\cscr L(h(s)) = \cscr L(W_L(s))$. In particular, \theo{theo:Main} ensures the existence of a constant $C_1$ such that $\E \$h\$ \le C_1$. By \lem{lem:estu}, there exists another constant $C_2$ such that $\E \|g\|_\infty < C_2$. Since the solution is Markovian, $g$ and $h$ are independent random variables and we have
\begin{equs}
\P\biggl(\|g\|_{\infty} < {2C_2\over \eps} \quad\text{and}\quad
\$h\$ < {2C_1\over \eps}\biggr) &= \P\Bigl(\|g\|_{\infty} < {2C_2\over \eps}\Bigr)\P\Bigl(\$h\$ < {2C_1\over \eps}\Bigr) \\
	&> (1-\eps/2)^2 > 1-\eps\;.
\end{equs}
From now on we assume that the above event is satisfied. Thus there is
a constant $C_3 \approx \CO(1/\eps)$ such that
\begin{equ}
\$\CM_{g,h} f\$ \le C_3 + k_2 T \$f\$^3\;.
\end{equ}
If we impose now $T < 1/(8k_2 C_3^2)$, we see that $\CM_{g,h}$ maps the
ball of radius $2C_3$ centered at $0$ into itself. If we also impose the condition
\begin{equ}
T < {1 \over k_3 (1+ 4C_3)^2}\;,
\end{equ}
we see that $\CM_{g,h}$ is a contraction on that ball. This, together with the uniqueness of the solutions of \eref{e:GL}, proves the claim. It moreover shows that the width $\eta$ of analyticity behaves asymptotically like $\eta \approx \CO(\eps)$.
\end{proof}
The above theorem tells us the probability for the solution to be
analytic in a strip at a fixed time. Another property of interest
is the behavior of the individual sample paths. We will show that any given sample path is always analytic with probability $1$. Recall that $\cscr F$ denotes the $\sigma$-field of the probability space underlying the cylindrical Wiener process.
\begin{proposition}
\label{prop:Anal}
There is an event $\Gamma \in \cscr F$ with $\P(\Gamma) = 1$ such that for every $\xi \in \L^\infty(\R)$, every $\omega \in \Gamma$, and every positive time $t>0$, there exists a strictly positive value $\eta(t)>0$ such that $u_\xi(t,\omega) \in \cscr A_\eta$.
\end{proposition}
\begin{proof}
Define for each integer $n$ the set $\Gamma_n$ as
\begin{equ}
\Gamma_n = \bigl\{w\in \Omega\,|\, W_L(\cdot,\omega) \in C([0,n],\cscr A_n)\bigr\}\;.
\end{equ}
We have $\P(\Gamma_n) = 1$ for all $n$ by \lem{lem:boundW}. By $\sigma$-completeness, $\Gamma = \bigcap_{n>0} \Gamma_n$ belongs to $\cscr F$ and $\P(\Gamma) = 1$. We claim that $\Gamma$ is the right event.

By the construction of $\Gamma$, the sample paths $u_\xi(\cdot,\omega)$ and $W_L(\cdot,\omega)$ are continuous and thus bounded on every finite time interval. Furthermore, $W_L(t,\omega) \in \cscr A_\eta$ for every time and every positive $\eta$. The claim now follows easily from the proof of \theo{theo:estAnal}.
\end{proof}
%
%
%
\section{Existence of an Invariant Measure}
\label{sec:inv}
%
%
%
We can now turn to the proof of \theo{theo:existM}. We first define the set of weight functions $\cscr W$ as the set of all functions $\rho : \R \to \R$ which satisfy
\begin{claim}
\item[\it a.] The function $\rho(x)$ is bounded, two times continuously differentiable and strictly positive.
\item[\it b.] For every $\eps>0$ there exists
$x_\eps > 0$ such that $|\rho(x)| \le \eps$ if $|x| \ge x_\eps$.
\item[\it c.] There exist constants $c_1$ and $c_2$ such that
\begin{equ}[e:proprho]
\Bigl|{\d_x \rho(x) \over \rho(x)}\Bigr|\le c_1\qquad\text{and}\qquad
\Bigl|{\d_x^2 \rho(x) \over \rho(x)}\Bigr|\le c_2\;,
\end{equ}
for all $x \in \R$.
\end{claim}
\begin{remark}
The meaning of the expression ``slowly decaying'' used in \theo{theo:existM} becomes clear from the following statement, the verification of which we leave to the reader.
For \emph{every} strictly positive decreasing sequence $\{x_n\}_{n=0}^\infty$ satisfying $\lim_{n \to \infty} x_n = 0$ and such that $x_{n}/x_{n+1}$ remains bounded, it is possible
to construct a function $\rho \in \cscr W$ such that $\rho(n) = x_{|n|}$ for every $n \in\Z$. In particular, $x_n$ may decay as slowly as
$1/\log(\log(\ldots\log(C+n)\ldots))$, but is not allowed to decay faster than exponentially.
\end{remark}
For every $\rho \in \cscr W$, we define the weighted norm
\begin{equ}
\|f\|_\rho = \|\rho f\|_\infty\;.
\end{equ}
We can now consider the topological vector space $\B_\rho$ which is equal as a set to $\B = C_u(\R)$, but endowed with the (slightly weaker) topology induced by the norm $\|\cdot\|_\rho$.
The space $\B_\rho$ is a metric space, but it is neither complete nor separable.
Since the topology of $\B_\rho$ is weaker than that of the original space $\B$, every $\B_\rho$-Borel set is also a $\B$-Borel set and every probability measure on $\B$ can be restricted to a probability measure on $\B_\rho$.
Let us show that we can define consistently a transition semigroup $\CP_{t,\rho}^*$ acting on and into the set of $\B_\rho$-Borel probability measures. We have
\begin{proposition}
\label{prop:trans}
For every $\rho \in \cscr W$, the transition semigroup $\CP_t^*$
associated to \eref{e:GL} can be extended to a transition semigroup
$\CP_{t,\rho}^*$ such that \eref{e:defAdj} holds for every $\B_\rho$-Borel set $\Gamma$. Furthermore, the transition semigroup $\CP_{t,\rho}^*$ is weak Feller.
\end{proposition}
In order to prove this proposition, we will show the Lipschitz continuous dependence of the solutions on the initial conditions in the new topology. For this, we need (see \app{app:Diss} for the definition of a dissipative mapping in a Banach space):
\begin{lemma}
\label{lem:dissRho}
The operator $\Delta$ is quasi dissipative with respect to the norm $\|\cdot\|_\rho$.
\end{lemma}
\begin{proof}
We have the equality
\begin{equ}
\rho\Delta u = \Delta(\rho u) - {\Delta \rho \over \rho}(\rho u) + 2{\nabla \rho \over \rho}\nabla(\rho u) - 2\Bigl|{\nabla\rho \over \rho}\Bigr|^2 (\rho u)\;.
\end{equ}
The claim follows from \eref{e:proprho} and the fact that $\Delta$ and $\nabla$ are dissipative operators with respect to $\|\cdot\|_\infty$.
\end{proof}
\begin{proof}[of \prop{prop:trans}]
Using \lem{lem:dissRho}, it is easy to check that the operator $L + F_t$ is, for all times and for a $\qua\in\R$, $\qua$-quasi dissipative with respect to the norm $\|\cdot\|_\rho$. 
This yields as in \lem{lem:detequ} the estimate
\begin{equ}
\|S_\xi^T(W) - S_\eta^T(W)\|_\rho \le e^{\qua T}\|\xi - \eta\|_\rho\;.
\end{equ}
Using this estimate, we can retrace the arguments exposed in the proof of \theo{theo:sol} to show that $\CP_{t,\rho}^*$ is well-defined and weak Feller.
\end{proof}
This construction is reminiscent of what was done in \cite{MS,FLS} to construct an attractor for the deterministic case. They also introduce a weighted topology on $\L^\infty(\R)$ to overcome the fact that the attractor of the deterministic Ginzburg-Landau equation is not compact.
Our result is the following.
\begin{theorem}
\label{theo:exist}
For every $\rho \in \cscr W$, there exists a $\B_\rho$-Borel probability measure $\mu_\rho$ which is invariant for the transition semigroup $\CP_{\rho,t}^*$.
\end{theorem}

The proof follows from a standard tightness argument. The main point is to notice that the unit ball of $\cscr A_\eta$ is compact in $\B_\rho$ for any weight function $\rho \in \cscr W$. We formulate this as a lemma.
\begin{lemma}
\label{lem:compact}
The unit ball of $\cscr A_\eta$ is a compact subset of $\B_\rho$ for every $\rho \in \cscr W$.
\end{lemma}
\begin{proof}
Since $\B_\rho$ is a metric space, compact sets coincide with sequentially compact sets \cite{Ko}. We use the latter characterization.
Choose a sequence $\CF = \{f_n\}_{n=1}^\infty$ of functions in $\cscr A_\eta$ with
$\$f_n\$_{\eta,\infty} \le 1$ for all $n$. It is a standard theorem of complex analysis \cite{Dieu} that if $\CD \subset \C$ is open and $\CF$ is a family of analytic functions uniformly bounded on $\CD$, then for every compact domain $K \subset \CD$ there is a subsequence of $\CF$ that converges uniformly on $K$ to an analytic limit.

We define the subsequences $\CF_n$ inductively by the following construction. First we choose $\CF_{-1} = \CF$. Then we consider the compact sets
$\CD_n = [-n,n]$ and we define $\CF_{n}$ as a subsequence of $\CF_{n-1}$ that converges uniformly on $\CD_n$. Call $\hat f_n$ the resulting limit function on $\CD_n$. We now define a global limit function $\hat f_\infty$ by $\hat f_\infty(x) = \hat f_n(x)$ if $x \in \CD_n$. This procedure is well-defined since different $\hat f_n$ must by construction coincide on the intersection of their domains.

It remains now to exhibit a subsequence of $\CF$ that converges to $\hat f_\infty$ in the topology of $\B_\rho$. For every $n\ge 1$, choose $g_n \in \CF_n$ such that $|g_n(z) - f_n(z)| < 1/n$ for $z \in \CD_n$. The $g_n$ form a subsequence of $\CF$. We have moreover
\begin{equ}
\|g_n - \hat f_\infty\|_\rho \le \|g_n - \hat f_N\|_\rho + \|\hat f_N - \hat f_\infty\|_\rho \le {\|\rho\|_\infty \over N} + 4\sup_{|x| \ge N} |\rho(x)|\;.
\end{equ}
By hypotheses {\it a.}~and {\it b.}~on $\rho$, this expression tends to $0$ as $N$ tends to $\infty$.
\end{proof}
\begin{remark}
By the compatibility of the various topologies with the linear structures, every bound\-ed closed subset of $\cscr A_\eta$ is compact as a subset of $\B_\rho$.
\end{remark}
\begin{proof}[of \theo{theo:exist}]
We choose an initial condition $\xi \in \L^\infty(\R)$ and consider the family of $\B_\rho$-Borel probability measures given by
\begin{equ}
\mu_t = {1 \over t}\int_0^t \CP^*_{\rho,t}(\delta_{\xi})\,dt\;.
\end{equ}
Fix now an arbitrary $\eps > 0$. By \theo{theo:estAnal} there exist $\eta, C, T > 0$ such that
$\mu_t(\CB(\cscr A_\eta,C)) > 1-\eps$ for every $t > T$. Since $\CB(\cscr A_\eta,C)$ is compact in $\B_\rho$ by \lem{lem:compact}, the family $\{\mu_t\}_{t>T}$ is tight and thus contains a weakly convergent subsequence by Prohorov's theorem. Denote by $\mu_\rho$ the limit measure.
Remember that a Borel probability measure on a metric space $M$ is uniquely determined by its values on $C_u(M)$ \cite{Bi}. The weak Feller property of $\CP_{t,\rho}^*$ is thus sufficient to retrace the proof of the Krylov-Bogoluboff existence theorem \cite{BK,ZDP}, which states that $\mu_\rho$ is invariant for $\CP^*_{\rho,t}$.
\end{proof}
%
%
%
\makeappendix{Dissipative Maps}
\label{app:Diss}
%
%
%
This appendix will first give a short caracterization of dissipative maps in Banach spaces. We will then prove a global existence theorem for the solutions of non-autonomous semilinear PDE's with a dissipative nonlinearity.
\begin{definition}
Given a Banach space $\B$ and a map $F:\CD(F)\subset \B \to \B$, one says \cite{ZDP1} that $F$ is \emph{dissipative} if 
\begin{equ}[e:defDiss]
\|x-y\|\le \bigl\|x-y - \alpha\bigl(F(x)-F(y)\bigr)\bigr\|\;,
\end{equ}
holds for every $x,y \in \CD(F)$ and every $\alpha > 0$. If there exists a $\qua \in \R$ such that $x \mapsto F(x) - \qua x$ is dissipative, we say that $F$ is $\qua$-quasi dissipative (or quasi dissipative for short).
\end{definition}
In the following, $u:(0,\infty) \to \B$ denotes a differentiable map. The function $\|u(\cdot)\|$ is of course continuous and its left-handed lower Dini derivative satisfies the inequality
\begin{equs}
D_-\|u(t)\| &= \liminf_{h \to 0^+} {\|u(t)\| - \|u(t-h)\| \over h} \\
&\le \liminf_{h \to 0^+}\biggl({\|u(t)\| - \|u(t)-h\dot u(t)\| \over h} 
 + {\|u(t-h) - u(t) + h\dot u(t)\| \over h}\biggr)\\
&=\liminf_{h \to 0^+}{\|u(t)\| - \|u(t)-h\dot u(t)\| \over h} \label{e:Dini}\;.
\end{equs}
This estimate allows to get easily very useful estimates on the norm of the solutions of dissipative differential equations. For example, if $\dot u(t) = F(u(t))$ holds for all times and $F$ is $\qua$-quasi dissipative, then the estimate
\begin{equ}[e:estDiss]
\|u(t)\| \le e^{\qua t}\bigl|\,\|u(0)\| - \|F(0)\|\,\bigr|+\|F(0)\|
\end{equ}
holds as a consequence of a standard theorem about differential inequalities \cite{Wa}.

We will now use standard techniques to prove a global existence theorem for the Cauchy problem
\def\dL{\overline{\CD(L)}}
\begin{equ}[e:Cauch]
\dot X_\xi(t) = L X_\xi(t) + F_t\bigl(X_\xi(t)\bigr)\;,\quad X_\xi(0) = \xi\;,
\end{equ}
and the associated integral equation
\begin{equ}[e:Int]
X_\xi(t) = e^{Lt}\xi + \int_0^te^{L(t-s)}F_s\bigl(X_\xi(s)\bigr)\,ds\;,
\end{equ}
in a Banach space $\B$. We do {\it not} require that the domain of $L$ be dense in $\B$. Let us denote by $\dL$ the Banach space obtained by closing the domain of $L$ in $\B$. Since, by assumption \ass{1} below, $L$ is chosen to be closed, we can equip $\CD(L)$ with the graph norm $\|x\|_L = \|x\| + \|Lx\|$ to obtain a Banach space. Our assumptions on $L$ and $F_t$ will be the following.
\begin{assum}{1}
The operator $L$ is sectorial in the sense that its resolvent set contains the complement of a sector in the complex plane and that its resolvent satisfies the usual bounds \cite[Def~2.0.1]{Lun}.
\end{assum}
This assumption implies \cite{Lun} that $L$ generates an analytic semigroup $S(t)$ which is strongly continuous on $\dL$ and maps $\B$ into $\CD(L^k)$ for any $k\ge 0$. Furthermore, a bound of the form $\|S(t)\| \le Me^{\Omega t}$ holds. We will assume without loss of generality that $M\le 1$ and $\Omega=0$. The latter assumption can be made since a constant can always be added to the nonlinear part. The former assumption is only made for convenience to simplify the notations. All the results also hold for $M > 1$. Another useful property of $S(t)$ is that there exists a constant $c$ such that $\|S(t)\xi\|_L \le c t^{-1}\|\xi\|$ for $\xi \in \B$ and $t>0$.
\begin{assum}{2}
There exist a positive time $T$ and a real constant $\qua$ such that the mapping $x \mapsto Lx + F_t(x)$ is $\qua$-quasi dissipative for all times $t\in[0,T]$.
\end{assum}
This assumption will ensure the existence of the solutions up to the time $T$, which may be infinite.
\begin{assum}{3}
The function $F_t$ is everywhere defined and there exist continuous increasing functions $a, \tilde a:\R_+ \to \R_+$ such that
\begin{equa}[e:defa]
\|F_t(x)\| &\le a(\|x\|)\;,\\
\|F_t(x) - F_t(y)\| &\le \|x-y\|\cdot \tilde a\bigl(\|x\| + \|y\|\bigr)\;,
\end{equa}
holds for every $x,y \in \B$ and for every $t \in [0,T]$.
\end{assum}
\begin{assum}{4}
The map $F_t$ maps $\CD(L)$ into $\CD(L)$ for all times and there exist continuous at most polynomially growing functions $b, \tilde b:\R_+ \to \R_+$ such that
\begin{equa}[e:defb]
\|F_t(x)\|_L &\le b(\|x\|_L)\;,\\
\|F_t(x) - F_t(y)\|_L &\le \|x-y\|_L\cdot \tilde b\bigl(\|x\|_L + \|y\|_L\bigr)\;,
\end{equa}
holds for every $x,y \in \CD(L)$ and for every $t \in [0,T]$.
\end{assum}
\begin{assum}{5}
The mapping $t \mapsto F_t(x)$ is continuous as a mapping $[0,T] \to \B$ for every $x\in \B$, and as a mapping $[0,T] \to \CD(L)$ for every $x \in \CD(L)$.
\end{assum}
These assumptions allow us to show the existence of the solutions of \eref{e:Cauch} in the mild sense for any initial condition $\xi \in \B$ and in the strict sense for $\xi \in \CD(L)$. Furthermore, we show that for any initial condition $\xi \in \B$, the solution lies in $\CD(L)$ after an infinitesimal amount of time. Similar results can be found in the literature (see \eg \cite{Lun,Hen} and references therein), but with slightly different assumptions. The present result has by no means the pretention to generality but is tailored to fit our needs. Since the proof is not excessively long, we give it here for the sake of completeness.
\begin{theorem}
\label{theo:Dissip}
Assume \ass{1}--\ass{5} hold and choose $\xi \in \B$. Then there exists a unique function $X_\xi:[0,T]\to \B$ solving \eref{e:Int} for $t\in[0,T]$. The solutions satisfy $\|X_\xi(t) - X_\eta(t)\| \le e^{\qua t}\|\xi-\eta\|$ for all times. Furthermore, $t \mapsto X_\xi(t)$ is differentiable for $t>0$, $X_\xi(t)\in \CD(L)$ and its derivative satisfies \eref{e:Cauch}.
\end{theorem}
\begin{proof}
\def\bB{\kern1mm\bar{\kern-1mm\cscr{B}}}
Assume first that the initial condition $\xi$ belongs to $\CD(L)$.
We denote by $\B_{L,T}$ the Banach space $C([0,T],\CD(L))$ with the usual sup norm. We show the local existence of a classical solution to \eref{e:Cauch} in $\B_{L,T}$ by a standard contraction argument.
Choose $T_0>0$ and define the map $\CM_\xi : \B_{L,T_0} \to \B_{L,T_0}$ by
\begin{equ}
\bigl(\CM_\xi f\bigr)(t) = S(t)\xi + \int_0^t S(t-s)F_s(f(s))\,ds\;.
\end{equ}
It is clear by \ass{1}, \ass{3}, \ass{4} and \ass{5} that $\CM_\xi$ is well-defined and that the bounds
\minilab{e:bounds}
\begin{equs}
\|\CM_\xi f\| &\le \|\xi\|_L + T_0 b(\|f\|)\;,\qquad \label{e:bound1}\\
\|\CM_\xi f - \CM_\xi g\| &\le T_0 \|f-g\|\cdot \tilde b(\|f\|+\|g\|)\;, \label{e:bound2} \\
\|\CM_\xi f - \CM_\zeta f\| &\le \|\xi-\zeta\|_L\;, \label{e:bound3}
\end{equs}
hold. It is clearly enough to take $T_0$ small enough, for example
\begin{equ}[e:boundT0]
T_0 < \min\biggl\{ {\|\xi\|_L\over b(2\|\xi\|_L)}\,,\;{1 \over \tilde b(4\|\xi\|_L)}\biggr\}\;,
\end{equ}
to find a contraction in the ball $\CB(\B_{L,T_0},2\|\xi\|_L)$. Thus $\CM_\xi$ possesses a unique fixed point $X_\xi$ in $\B_{L,T_0}$. By \cite[Lem.~4.1.6]{Lun}, $X_\xi$ is strongly differentiable in $\B$ and its derivative satisfies \eref{e:Cauch}.

Using \eref{e:Dini} and \ass{2}, we see immediately that for any $\xi,\zeta \in \CD(L)$ and $t>0$ such that the strong solutions $X_\zeta$ and $X_\xi$ exist up to time $t$, the estimates
\begin{equa}[e:estSol]
\|X_\xi(t)\| &\le \bigl|\|\xi\| - a(0)\bigr| e^{\qua t} + a(0)\;, \\
\|X_\zeta(t) - X_\xi(t)\| &\le e^{\qua t}\|\zeta-\xi\| \;,
\end{equa}
hold.
The global existence of the solution now follows by iterating the above arguments, using \eref{e:estSol} to ensure the non-explosion of the solutions. We leave it to the reader to verify that one can indeed continue the solutions up to the time $T$.

We next now show that for any initial condition $\xi\in\B$, the solution of \eref{e:Int} exists locally and lies in $\CD(L)$ for positive times. We define $\CM_\xi$ as above, but replace the space $\B_{L,T_0}$ by the larger space $\bB_{L,T_0}$ given by the measurable functions $f : (0,T_0] \to \CD(L)$ with finite norm
\begin{equ}
\$f\$ = \sup_{t\in(0,T_0]}\|t f(t)\|_L + \sup_{t\in(0,T_0]}\|f(t)\|\;.
\end{equ}
We first show that $\CM_\xi$ is well-defined on $\bB_{L,T_0}$. Choose $f\in \bB_{L,T_0}$. It is easy to check that, by \ass{3}, $\bigl\|\bigl(\CM_\xi f\bigr)(t)\bigr\|\le \|\eta\| + T_0a(\$f\$)$. By \ass{4}, we can choose $n$ such that $b$ and $\tilde b$ grow slower than $(1+x)^n$. We also choose an exponent $N > n$ and choose $T_0 < 1$.
We have, by the remark following \ass{1}, the estimate
\begin{equs}
\bigl\|t \bigl(\CM_\xi f\bigr)(t)\bigr\|_L &\le \|tS(t)\xi\|_L + \int_0^{t-t^N}\Bigl\|t S(t-s) F_s(f(s))\Bigr\|_L\,ds \\
	&\qquad + \int_{t-t^N}^t\Bigl\|t S(t-s) F_s(f(s))\Bigr\|_L\,ds\\
	&\le c\|\xi\| + \int_0^{t-t^N}{ct\over t-s} a(\|f(s)\|)\,ds + \int_{t-t^N}^{t}t b(\|f(s)\|_L)\,ds\\
	&\le c\|\xi\| +  C_1 t\ln(t)a(\$f\$) + C_2 t^{N+1}\Bigl(1+{\$f\$\over t}\Bigr)^n\;.
\end{equs}
A similar estimate holds for $\$\CM_\xi f - \CM_\xi g\$$.
Since $N>n$, there exists a function $\chi$ such that estimates of the type
\begin{equ}
\$\CM_\xi f\$ \le \sqrt{T_0} \chi(\$f\$)\quad\text{and}\quad
\$\CM_\xi f - \CM_\xi g\$ \le \sqrt{T_0} \$f-g\$ \chi(\$f\$ + \$g\$)
\end{equ}
hold. It follows that $T_0$ can be chosen sufficiently small to make $\CM_\xi$ a contraction on some ball of $\bB_{L,T_0}$, and so the fixed point of $\CM_\xi$ takes its values in $\CD(L)$.

In order to complete the proof of the theorem, it remains to show that \eref{e:estSol} holds for arbitrary intial conditions. We again consider the same mapping $\CM_\xi$, but this time on $C_b((0,T_0],\B)$.
It is straightforward to check, using the assumptions, that bounds similar to \eref{e:bounds}, but with $\|\cdot\|_L$ replaced by $\|\cdot\|$ and $b$, $\tilde b$ replaced by $a$, $\tilde a$ hold.
We notice that, by \eref{e:bound1}, we can, for arbitrary $\eps > 0$, choose $\delta$ so small that $\|u_\eta(\delta)\| \le (1+\eps)\|\eta\|$. Since $u(\delta) \in \CD(L)$, this gives the estimate $\|u_\eta(t)\| \le |(1+\eps)\|\eta\|-a(0)|e^{\qua(t-\delta)}+a(0)$, holding for every $\eps>0$. By using \eref{e:bound2} and a similar argument, we can show that $\|u_\eta(t) - u_\xi(t)\| \le e^{\qua(t-\delta)}(1+\eps)\|\eta-\xi\|$ holds and thus \eref{e:estSol} is true for $\eta, \xi\in\B$.
\end{proof}
\begin{acknowledge}
We would like to thank Marius Mantoiou, Radu Purice, Guillaume van Baalen and Peter Wittwer for helpful discussions. This work was partially supported by the Fonds National Suisse.
\end{acknowledge}
\bibliographystyle{myalph}
\markboth{\sc \refname}{\sc \refname}
\bibliography{refs}

\def\Rom#1{\uppercase\expandafter{\romannumeral #1}}
\providecommand{\bysame}{\leavevmode\hbox to3em{\hrulefill}\thinspace}
\begin{thebibliography}{DPZ92b}

\bibitem[Bil68]{Bi}
P.~Billingsley, \emph{Convergence of Probability Measures}, {John Wiley \&
  Sons}, 1968.

\bibitem[BK37]{BK}
N.~N. Bogoluboff and N.~M. Kriloff, \emph{La Th\'eorie G\'en\'erale de la
  Mesure dans son Application \`a l'Etude des Syst\`emes Dynamiques de la
  M\'ecanique Non-Lin\'eaire}, Ann. of Math. \textbf{38} (1937), 65--113.

\bibitem[BKL00]{BC}
J.~Bricmont, A.~Kupiainen, and R.~Lefevre, \emph{Ergodicity of the 2D
  Navier-Stokes Equations with Random Forcing}, Preprint, 2000.

\bibitem[Col94]{PC}
P.~Collet, \emph{Non Linear Parabolic Evolutions in Unbounded Domains}, NATO
  Adv. Sci. Inst. Ser. C Math. Phys. Sci \textbf{437} (1994), 97--104.

\bibitem[Die68]{Dieu}
J.~Dieudonn{\'e}, \emph{Foundations of Modern Analysis}, 7th ed., {Aca\-demic
  Press}, New York and London, 1968.

\bibitem[DPZ92a]{DZA}
G.~Da~Prato and J.~Zabczyk, \emph{Non-explosion, Boundedness and Ergodicity for
  Stochastic Semilinear Equations}, {J.~Diff.~Equ.} \textbf{98} (1992),
  181--195.

\bibitem[DPZ92b]{ZDP1}
G.~Da~Prato and J.~Zabczyk, \emph{Stochastic Equations in Infinite Dimensions},
  {University Press}, {Cambridge}, 1992.

\bibitem[DPZ96]{ZDP}
G.~Da~Prato and J.~Zabczyk, \emph{Ergodicity for Infinite Dimensional Systems},
  London Mathematical Society Lecture Note Series, vol. 229, {University
  Press}, {Cambridge}, 1996.

\bibitem[EPR99]{EPR2}
J.-P. Eckmann, C.-A. Pillet, and L.~R{ey-Bellet}, \emph{Entropy Production in
  Non-Linear, Thermally Driven Hamiltonian Systems}, {J.~Stat.~Phys.}
  \textbf{95} (1999), 305--331.

\bibitem[FLS96]{FLS}
E.~Feireisl, P.~Lauren{\c c}ot, and F.~Simondon, \emph{Global Attractors for
  Degenerate Parabolic Equations on Unbounded Domains}, {J.~Diff.~Equ.}
  \textbf{129} (1996), 239--261.

\bibitem[Hen81]{Hen}
D.~Henry, \emph{Geometric Theory of Semilinear Parabolic Equations}, Lecture
  Notes in Mathematics, vol. 840, {Springer}, {New York}, 1981.

\bibitem[JLM85]{JLM}
G.~Jona-Lasinio and P.~K. Mitter, \emph{On the Stochastic Quantization of Field
  Theory}, {Comm.~Math.~Phys.} \textbf{101} (1985), 406--436.

\bibitem[K{\"o}t83]{Ko}
G.~K{\"o}the, \emph{Topological Vector Spaces {\Rom 1--\Rom 2}}, {Springer},
  {New York}, 1983.

\bibitem[Lun95]{Lun}
A.~Lunardi, \emph{Analytic Semigroups and Optimal Regularity in Parabolic
  Problems}, {Birkh{\"a}user}, {Basel}, 1995.

\bibitem[MS95]{MS}
A.~Mielke and G.~Schneider, \emph{Attractors for Modulation Equations on
  Unbounded Domains -- Existence and Comparaison}, {Nonlinearity} \textbf{8}
  (1995), 743--768.

\bibitem[RS80]{RS}
M.~Reed and B.~Simon, \emph{Methods of Modern Mathematical Physics {\Rom
  1--\Rom 4}}, {Aca\-demic Press}, {San Diego, California}, 1980.

\bibitem[Wal64]{Wa}
W.~Walter, \emph{Differential- und Integralungleichungen}, Springer Tracts in
  Natural Philosophy, vol.~2, {Springer}, {New York}, 1964.

\end{thebibliography}

\end{document}